\documentstyle[12pt,psfig,epsfig]{article}

\newcommand{\resection}[1]{\setcounter{equation}{0}\section{#1}}

\newcommand{\EQ}{\begin{equation}}
\newcommand{\EN}{\end{equation}}
\newcommand{\bea}{\begin{eqnarray}}
\newcommand{\eea}{\end{eqnarray}}

\newcommand{\th}{\theta}

\newcommand{\goto}{\rightarrow}

\newfont{\twelvemsb}{msbm10 scaled\magstep1}
\newfont{\eightmsb}{msbm8}
\newfam\msbfam
\textfont\msbfam=\twelvemsb
\scriptfont\msbfam=\eightmsb
\catcode`\@=11
\def\Bbb{\ifmmode\let\next\Bbb@\else
  \def\next{\errmessage{Use \string\Bbb\space only in math mode}}\fi\next}
\def\Bbb@#1{{\fam\msbfam{{#1}}}}

\def\EQ{\begin{equation}}
\def\EN{\end{equation}}
\def\ch{{\rm ch}}
\def\sh{ {\rm sh}}

\def\tilde{\widetilde}

\def\*{\star}
\def\[{\left[}
\def\]{\right]}
\def\({\left(}		
\def\){\right)}

\def\frac#1#2{{#1 \over #2}}
\def\inv#1{{1 \over #1}}

\def\d{\partial}

\def\2pi{\hbox{$2\pi i$}}

\def\dsl{\raise.15ex\hbox{/}\kern-.57em\partial}
\def\Dsl{\,\raise.15ex\hbox{/}\mkern-.13.5mu D}
\def\th{\theta}

\def\vep{\varepsilon}

		\def\CO{{\cal O}}

\def\2pi{\hbox{$2\pi i$}}

\def\dsl{\raise.15ex\hbox{/}\kern-.57em\partial}
\def\Dsl{\,\raise.15ex\hbox{/}\mkern-.13.5mu D}
\font\numbers=cmss12
\font\upright=cmu10 scaled\magstep1
\def\stroke{\vrule height8pt width0.4pt depth-0.1pt}
\def\topfleck{\vrule height8pt width0.5pt depth-5.9pt}
\def\botfleck{\vrule height2pt width0.5pt depth0.1pt}
\def\Zmath{\vcenter{\hbox{\numbers\rlap{\rlap{Z}\kern
0.8pt\topfleck}\kern 2.2pt
                   \rlap Z\kern 6pt\botfleck\kern 1pt}}}
\def\Qmath{\vcenter{\hbox{\upright\rlap{\rlap{Q}\kern
                   3.8pt\stroke}\phantom{Q}}}}
\def\Nmath{\vcenter{\hbox{\upright\rlap{I}\kern 1.7pt N}}}
\def\Cmath{\vcenter{\hbox{\upright\rlap{\rlap{C}\kern
                   3.8pt\stroke}\phantom{C}}}}
\def\Rmath{\vcenter{\hbox{\upright\rlap{I}\kern 1.7pt R}}}
\def\Z{\ifmmode\Zmath\else$\Zmath$\fi}
\def\Q{\ifmmode\Qmath\else$\Qmath$\fi}
\def\N{\ifmmode\Nmath\else$\Nmath$\fi}
\def\C{\ifmmode\Cmath\else$\Cmath$\fi}
\def\R{\ifmmode\Rmath\else$\Rmath$\fi}

\begin{document}
\setcounter{page}{0}
\topmargin 0pt
\oddsidemargin 5mm
\renewcommand{\thefootnote}{\arabic{footnote}}
\newpage
\setcounter{page}{0}
\begin{titlepage}
\begin{flushright}
SISSA/27/2001/FM
\end{flushright}
\vspace{0.5cm}
\begin{center}
{\large {\bf On the Finite Temperature Formalism in \\
Integrable Quantum Field Theories}} \\
\vspace{1.8cm}
{\large G. Mussardo$^{a,b}$} \\
\vspace{0.5cm}
{\em $^{a}$International School for Advanced Studies, Trieste, Italy}\\
\vspace{0.3cm}
{\em $^{b}$Istituto Nazionale di Fisica Nucleare, Sezione di Trieste}\\
\end{center}
\vspace{1.2cm}

\renewcommand{\thefootnote}{\arabic{footnote}}
\setcounter{footnote}{0}

\begin{abstract}
\noindent
Two different theoretical formulations of the finite temperature
effects have been recently proposed for integrable field theories.
In order to decide which of them is the correct one,
we perform for a particular model an explicit check of their predictions
for the one--point function of the trace of the stress--energy tensor,
a quantity which can be independently determined by the Thermodynamical
Bethe Ansatz.
\end{abstract}

\vspace{.5cm}

\end{titlepage}

\newpage

\resection{Introduction}
Finite temperature correlation functions are important quantities
for many applications of both theoretical and experimental interest
(see, for instance \cite{Tsvelik}). A special class of quantum
field theories is provided by the two--dimensional integrable
models, which can be exactly solved by means of bootstrap
methods \cite{Zam,Musrep,Smirnov,ZamTBA}. For these models,
two different formulations of finite temperature effects
have been recently discussed in the literature: the first
is due to LeClair and Mussardo \cite{LM}, the second has been
proposed by Delfino \cite{Del}. Although the two formalisms
coincide if applied to the trivial cases of free quantum field
theories, however they drastically differ once used to deal with
interacting theories. To determine which of the two is the correct
one we decide to compare their predictions versus a quantity
which can be independently determined. This is the case of
the finite temperature one--point function of the trace of the
stress--energy tensor which can be computed by the
Thermodynamical Bethe Ansatz (TBA) \cite{ZamTBA}. As we will show
below, the proposal by LeClair and Mussardo exactly matches the
low--temperature expansion of this quantity whereas the proposal
by Delfino fails at order ${\cal O}(e^{-3 mr})$. Before presenting
the explicit calculations, let us briefly discuss the main features
of the two different finite temperature formalisms.

\resection{LeClair--Mussardo Formalism}
This formalism, discussed in \cite{LM}, combines together physical
principles coming from two different areas: the Thermodynamical
Bethe Ansatz and the Form Factor Approach. It originates from an
interpretation of the expression of the free energy -- as determined
by the TBA --, in terms of quasi--particle excitations with respect
to a thermal ground state. In order to clarify this statement, it
is useful to summarise the TBA approach. We assume for simplicity
that the spectrum of the integrable theory consists of a single
particle $A$ with mass $m$ and an exact $S$--matrix $S(\theta)$. In
the following we consider the case $S(0) = -1$, which gives rise
to the fermionic TBA equations. We define
\begin{equation}
\label{2.8}
\sigma (\theta) = -i \log S (\theta) , ~~~~~
\phi (\theta) = -i \frac{d}{d\theta} \log S (\theta) \,\,\,.
\end{equation}
The partition function at a finite temperature $T$ and on a volume
$L$ (for $L \rightarrow \infty$) is determined by means of the
Thermodynamical Bethe Ansatz equations as follows \cite{ZamTBA}.
In a box of large volume $L$, $0<x<L$, with periodic boundary
conditions, the quantization condition of the momenta is given by
$e^{ik(\th_i ) L } \prod_{j\neq i}  S(\th_i - \th_j ) = 1$,
i.e.
\begin{equation}
\label{2.9}
m L \,\sh \th_i + \sum_{j\neq i} \sigma (\th_i - \th_j ) = 2\pi n_i
\,\,\, ,
\end{equation}
where $n_i$ are integers.  Introducing a density of occupied states
per unit volume $\rho_1 (\theta)$ as well as a density of levels
$\rho (\theta) $, in the thermodynamic limit eq.\,(\ref{2.9}) becomes
\begin{equation}
\label{2.10}
2\pi \rho = e + 2\pi \phi * \rho_1 \,\,\, ,
\end{equation}
where $e = m \cosh\theta$ and $(f*g)(\th ) =
\int_{-\infty}^\infty d\th' f(\th - \th') g(\th') /2\pi $.
Defining the pseudo-energy $\vep (\th)$ as
\begin{equation}
\label{2.11}
\frac{\rho_1}{\rho} = \inv{1+ e^{\vep} } \,\,\, ,
\end{equation}
the minimization of the free-energy with respect to the densities of
states leads to the integral equation
\begin{equation}
\label{2.12}
\vep = e R - \phi * \log ( 1 + e^{-\vep}  ) \,\,\, ,
\end{equation}
and the partition function is then given by
\begin{equation}
\label{2.13}
Z (L, R) = \exp\left[ mL
\int \frac{d\th}{2\pi} \ch \th ~ \log
\( 1 + e^{-\vep (\th )} \)\right] \,\,\, .
\end{equation}
As shown in \cite{LM}, the interesting point is now that the above
partition function can be interpreted as the one of a gas of
fermionic particles but with energy given by $\vep(\th)/R$. Namely,
there is a one--to--one correspondance between the above expression
(\ref{2.13}) and the partition function computed according to the
following thermal sum
\begin{equation}
\label{2.14}
Z(L,R) = \sum_{n=0}^\infty \inv{n!}
\int \frac{d\th_1}{2\pi} \cdots \frac{d\th_n}{2\pi}
~ \langle \th_n \cdots \th_1 | \th_1 \cdots \th_n \rangle ~
\prod_{i=1}^n e^{-\vep (\th_i ) } \,\,\, ,
\end{equation}
where the scalar products of the states are computed by applying the
standard free fermionic rules. The above equality implies that all
physical properties of the system can be extracted by employing the
quasi-particle excitations above the TBA thermal ground state. Since
this differs from the usual (zero--temperature) ground state, it is not
surprising that its excitations do not satisfy the standard dispersion
relations $e = m \cosh\beta$, $p = m \sinh\beta$, rather they have
dressed energy $\tilde{e} = \vep(\th) /R$ and
dressed momentum $\tilde{k} (\th)$:
\begin{equation}
\label{2.17}
\tilde{e} (\th ) = \vep (\th )/R \, , ~~~~~~
\tilde{k} (\th ) = k(\th) + 2\pi (\sigma * \rho_1 ) (\th ) \,\,\,.
\end{equation}
In this contest, the rapidity $\th$ plays the role of a variable which
simply parameterises the dispersion relation of the quasi--particle
excitations and their $S$--matrix, which is assumed to coincide with
the original $S(\theta_i-\theta_j)$.

The TBA allows us to compute the finite temperature one--point
function of the trace of the stress--energy tensor $T_{\mu}^{\mu}$
\cite{ZamTBA}. In fact, we have
\EQ
\langle T_\mu^\mu \rangle_R - (T_\mu^\mu)_0
= \frac{2\pi}{R}  \frac{d}{dR}  [RE(R)] \,\,\, ,
\EN
where $E(R) = -\log Z /L$.  This can be also expressed as
\begin{equation}
\label{2.20}
\langle T_\mu^\mu \rangle_R - (T_\mu^\mu)_0 = m \int d\th
\frac{e^{-\vep}}{1+ e^{-\vep} }
\(  \d_R \vep \,  \ch \th  - \inv{R}  \d_\th \vep \, \sh \th \)
\,\,\, ,
\end{equation}
where the functions $\d_R \vep$ and $\d_\th \vep$ satisfy linear integral
equations which can be easily solved. The final result reads
\begin{equation}
\label{2.25}
\langle T^\mu_\mu \rangle_R - (T_\mu^\mu)_0 = 2\pi m^2
\( \sum_{n=1}^\infty \int
\[ \prod_{i=1}^n   \frac{d\th_i}{2\pi} f (\th_i )
e^{-\vep(\theta_i)}\]
\phi (\th_{12} ) \cdots \phi(\th_{n-1, n} )\,\ch (\th_{1n} )
\) \,\,\, ,
\label{seriestrace}
\end{equation}
where
\begin{equation}
\label{2.7}
f(\theta ) = \inv{ 1 + e^{-\vep (\th)} } \,\,\, .
\end{equation}

Let us consider now the calculation of the finite temperature
one--point functions (the only ones which we consider in this paper).
According to LeClair and Mussardo, this correlator is given by
\begin{equation}
\label{2.6}
\langle {\cal O} (x,t) \rangle_R
= \sum_{n=0}^\infty \inv{n!} \frac{1}{(2\pi)^n}
\int \left[\prod_{i=1}^n d\th_i f(\th_i ) e^{-\vep (\th_i) }
\right]
\langle \th_n \cdots \th_1 | {\cal O} (0) | \th_1 \cdots \th_n
\rangle_{\rm conn}
\, ,
\end{equation}
where the connected Form Factor of the operator ${\cal O}$ is defined
as
\begin{equation}
\label{2.5}
\langle \th_n \cdots \th_1 | \CO  | \th'_1 \cdots \th'_m
\rangle_{\rm conn}
\equiv {\large FP}
\( \lim_{\eta_i \to 0 }
\langle 0 | \CO  | \th_n +i\pi + i \eta_n , \ldots , \th_1 +i\pi +
i \eta_1,\theta_1,\ldots,\theta_n \rangle \) \,\,\,
\end{equation}
$FP$ in front of the above expression means taking its {\em finite part},
i.e. terms proportional to $(1/\eta_i)^p$, where $p$ is some positive
power, and also terms proportional to $\eta_i / \eta_j $, $i\neq j$ are
discarded in taking the limit. With this prescription the resulting
expression is an universal quantity, i.e. independent of the way in which
the above limits are taken.

It is easy to see that within this formalism, the finite temperature
one--point function of the trace of the stress--energy tensor
exactly coincides with its expression provided by the TBA,
eq.\,(\ref{2.25}). In fact, the connected matrix elements of
this operators are given by
\begin{eqnarray}
\label{2.18}
\langle \th | T_\mu^\mu | \th \rangle_{\rm conn} &=& 2\pi m^2 \,\,\, ;
\\ \nonumber
\langle \th_2 , \th_1 | T_\mu^\mu | \th_1 , \th_2 \rangle_{\rm conn}
&=& 4\pi m^2 \phi(\th_1 - \th_2 ) \ch (\th_1 - \th_2 ) \,\,\, ,
\end{eqnarray}
and by an inductive application of the form factor residue equations
\begin{eqnarray}
\label{2.19}
\langle \th_n \cdots \th_1 | T_\mu^\mu | \th_1 \cdots \th_n \rangle_{\rm
conn} & = & 2\pi m^2 \, \phi (\th_{12} ) \phi (\th_{23}) \,\,\, ,
\cdots \phi(\th_{n-1,n})\,\ch (\th_{1n} )  \\
& & ~~~~~~+{\rm  permutations} \nonumber
\end{eqnarray}
where $\th_{ij} = \th_i - \th_j $. Once inserted into
eq.\,(\ref{2.6}), the above series coincides with the one
of eq.\,(\ref{2.25}).

In conclusion, the formalism by LeClair and Mussardo predicts,
at least for the particular thermal one--point function of
$T_{\mu}^{\mu}$, an exact matching with the expression determined
by the TBA.

\resection{Delfino's Formalism}
This formalism, discussed in \cite{Del}, only employs the
Form Factor Approach. The finite temperature effects are
taken into account by defining the theory on a cylinder
infinitely extended in the space direction and a width $R = 1/T$
in the other direction. The particles entering the thermal sum
are the asymptotic states satisfying the standard dispersion relations
$e = m \cosh\beta$, $p = m \sinh\beta$ and the contribution of
the $n$--particle asymptotic state to $Tr[{\cal O} e^{-H R}]$ is
given by
\begin{equation}
d_n^{\cal O}(R) =
\frac{1}{n!} \frac{1}{(2 \pi)^n}
\int d\theta_1\ldots d\theta_n
F_{n,n}^{\cal O}(\theta_n,\ldots,\theta_1 |
\theta_1,\ldots,\theta_n) e^{-E_n R} \,\,\,
\label{ncontr}
\end{equation}
with $E_n = m \sum_{i=1}^n \cosh\theta_i$ and
$$
F^{\cal O}_{m,n}(\theta_m',\ldots,\theta_1' |
\theta_1,\ldots,\theta_n) =
\langle \theta_m',\ldots,\theta_1' | {\cal O} |
\theta_1,\ldots,\theta_n\rangle \,\,\,.
$$
 Define
\begin{equation}
d^{\cal O}(R) = \sum_{n=0}^{\infty} d_n^{\cal O}(R) \,\,\, ,
\label{ratio}
\end{equation}
and normalise the thermal sum with respect the identity operator
$I$
\begin{equation}
\langle {\cal O} \rangle_{R} = \frac{d^{\cal O}(R)}{d^I(R)} \,\,\,.
\label{cancel}
\end{equation}
In his paper \cite{Del}, Delfino considered for the finite part
of the Form Factors entering eq.\,(\ref{ncontr}) the {\em symmetric
limit}
\begin{equation}
{\cal F}_{2n}^{\cal O}(\theta_1,\ldots,\theta_n) =
\lim_{\eta \goto 0}
F^{\cal O}_{0,2n}(\theta_1 + i\pi + i\eta,
\ldots,\theta_1+i\pi +i\eta,\theta_1,\ldots,
\theta_n) \,\,\,,
\label{symmetric}
\end{equation}
and he also showed that the singular disconned parts of the Form Factors
of the local operator ${\cal O}$ only enter through the constant factor
$S(0)$. All other singular terms cancel in the ratio (\ref{cancel}).
Finally, he proposed for the finite temperature one--point function
the expression
\begin{equation}
\langle {\cal O} \rangle_R =
\sum_{n=0}^{\infty} \frac{1}{n!} \frac{1}{(2 \pi)^n}
\int \left[\prod_{i=1}^n d\theta_i g(\theta_i,R) e^{-m R \cosh\theta_i}
\right] {\cal F}^{\cal O}_{2n}(\theta_1,\ldots,\theta_n)
\,\,\,,
\label{onedel}
\end{equation}
where
\begin{equation}
g(\theta,R) = \frac{1}{1 - S(0) e^{-m R \cosh\theta}} \,\,\,.
\label{functg}
\end{equation}
The above formula has to be contrasted with the one given by
eq.\,(\ref{2.6}).

\resection{Main Differences and Open Problems}
There are two main differences between the two formalisms:
\begin{itemize}
\item LeClair--Mussardo formalism employs the quasi--particle
excitations with respect to the thermal vacuum and therefore the
pseudo--energy $\vep(\theta)$, solution of the integral equation
(\ref{2.10}), whereas Delfino's formalism employs the standard
asymptotic particles at zero--temperature with energy $e=m\cosh\theta$
and momentum $p = m\sinh\theta$. These different choices of
excitations seem somehow related to the boundary conditions
adopted by the two formalisms along the space direction, i.e.
in the LeClair--Mussardo approach one considers a box of
large volume $L$, with periodic b.c., in the limit $L \goto \infty$,
whereas in the Delfino approach one directly considers the infinitely
extended line. Notice, however, that there is no dependence on $L$
in the final expressions (\ref{2.6}) and (\ref{onedel}) and therefore
it is not {\em a--priori} clear the role played by the boundary
conditions in thermal effects and which of the two is the appropriate one.
\item  the Form Factors entering equation (\ref{2.6}) are computed
according to the prescription given by eq.\,(\ref{2.5}) whereas those
entering equation (\ref{onedel}) are computed according to the symmetric
limit (\ref{symmetric}). The two different prescriptions
for the finite part of the Form Factors produce, of course,
two different results. In the case of the trace of the
stress--energy tensor, for instance, there is already a difference
for the two--particle Form Factor entering the thermal sum:
by using the symmetric limit, in fact we have
\begin{equation}
\langle \theta_2,\theta_1 | T_{\mu}^{\mu} |
\theta_1,\theta_2 \rangle = 8 \pi m^2 \phi(\theta_1-\theta_2)
\cosh^2\frac{\theta_1-\theta_2}{2} \,\,\,,
\label{2symm}
\end{equation}
to be contrasted with eq.\,(\ref{2.18}), obtained by using
the other prescription.
\end{itemize}
It is therefore evident that the two formulas, eq.\,(\ref{2.6})
and eq.\,(\ref{onedel}), proposed for the one--point function
at finite temperature, deeply differ in their physical
justifications and in their technical details. To judge which
of the two is the correct one it seems necessary to reach a better
understanding of the physical principles ruling the thermal
effects in quantum field theories. Given the present ignorance
about these principles, it is therefore difficult to decide
{\em a--priori} in favour of one or the other of the two
formulations and the best thing one can do is to to perform some
checks. Those already done and discussed in the literature
are unfortunately inconclusive. Lukyanov \cite{Lukyanov}, for
instance, computed the thermal one--point functions of the
vertex operators in the Sinh--Gordon model by performing the
path integral of the model and he showed that these quantities
coincide with the ones computed in the formalism by LeClair--Mussardo.
Unfortunately, the perturbative order at which he performed the
computation does not permit to decide about their general validity.
On the other hand, Delfino \cite{Del} showed that
his formalism is able to reproduce the one--point function of
$T_{\mu}^{\mu}$ up to the two--particle contributions but
unfortunately he did not prove the complete equivalence of his
formula with the TBA expression.

Given the present unsatisfactory status about the validity of the
two formalisms it is highly desiderable to perform additional checks,
in particular by comparing their predictions against a quantity
which can be determined by an independent method. These
considerations naturally select the one--point function of the
trace of the stress--energy tensor as a check quantity for the
two formulas, since its expression (\ref{2.25}) is independently
determined by the TBA. Hence, we have to see whether or not Delfino's
formula reproduces the TBA result, not only up to the two--particle
contribution, but also to higher orders (as shown above, the formula
by LeClair--Mussardo coincides with the formula of the TBA). We have
then two possibilities: (i) the formula proposed by Delfino is unable to
reproduce the TBA result at higher orders; (ii) the formula proposed
by Delfino reproduces the TBA result, alias it is just a different
organization of the terms entering both the thermal sum and the
integral equations of the TBA. In the first case, the failure
of this check is already enough to decide about the general validity
of the thermal expressions proposed by Delfino. In the second
case, there would be still open the problem which of the
two formalisms is the correct one, since their coincidence for
the particular case of the stress--energy tensor is not expected
to occur for other operators. Luckly enough, it is the first
possibility that happens. To show the discrepancy of Delfino's
formula with the TBA, we compare the thermal expression of the
stress--energy tensor of a particular model which can be
analytically solved.

\resection{A Simplified Model}
The main technical difficulty for comparing Delfino's expression
of $\langle T_{\mu}^{\mu}\rangle_R$ with the analogous expression
coming from the TBA consists in solving the integral equation
(\ref{2.10}). We can simplify this step by taking a local kernel,
i.e. we consider an integrable model for which
\begin{equation}
\phi(\theta_1-\theta_2) = 2\pi \delta(\theta_1-\theta_2) \,\,\,.
\label{delta}
\end{equation}
For the associate $S$--matrix we have
\begin{equation}
S(\theta) = \left\{
\begin{array}{ll}
1 & \mbox{if $\theta \neq 0$}\,\,\,; \\
-1 & \mbox{if $\theta = 0$} \,\,\,.
\end{array}
\right.
\end{equation}
The integrable model defined in this way may be regarded as the
limit $g \goto 0$ of the Sinh--Gordon model. In fact, with the notation
of ref.\,\cite{KM}, the $S$--matrix of the Sinh--Gordon model is given by
\begin{equation}
S_{Sh}(\theta) = \frac{\sinh\theta - i \sin\frac{\pi B(g)}{2}}
{\sinh\theta + i \sin\frac{\pi B(g)}{2}}\,\,\,,
\label{SSh}
\end{equation}
with $B(g) = \frac{2 g^2}{8\pi + g^2}$. It is convenient to define
$B(g) \equiv 2 \alpha$. For the corresponding kernel we have
\begin{equation}
\phi_{Sh}(\theta) = \frac{2 \sin\pi\alpha \cosh\theta}{\sinh^2\theta +
\sin^2\pi\alpha} \,\,\,,
\label{rdelta}
\end{equation}
and in the limit $\alpha \goto 0$ we have
\begin{equation}
\lim_{\alpha\goto 0} \phi_{Sh}(\theta) = 2 \pi \delta(\theta) \,\,\,.
\label{distribution}
\end{equation}
By using the kernel (\ref{delta}), the integral equation (\ref{2.10})
becomes
\begin{equation}
\vep(\theta) = mR \cosh\theta - \ln(1 + e^{-\vep(\theta)}) \,\,\,,
\label{tbaeq}
\end{equation}
whose solution is given by
\begin{equation}
\vep(\theta) = \ln\left(e^{mR \cosh\theta} -1\right)\,\,\,.
\label{eps}
\end{equation}
Hence
\begin{equation}
f(\theta) \,e^{-\vep(\theta)} = \frac{e^{-\vep}}{1+e^{-\vep}} =
e^{-mR \cosh\theta} \,\,\,,
\label{therfac}
\end{equation}
and inserting into the TBA formula (\ref{2.25}), we have
\begin{equation}
\langle T^\mu_\mu \rangle_R - (T_\mu^\mu)_0 = 2\pi m^2
\int_{-\infty}^{+\infty}\frac{d\theta}{2\pi}
\left[e^{-m R \cosh\theta} + e^{-2 mR \cosh\theta} +
e^{-3 mR \cosh\theta} + \cdots\right] \,\,\,.
\label{geometrical}
\end{equation}
For the purpose of comparing with Delfino's prediction, it is
convenient to leave explicitly the $n$--particle contributions
to the thermal average, although it is evident that the above
series can summed to
\begin{equation}
\langle T^\mu_\mu \rangle_R - (T_\mu^\mu)_0 = 2\pi m^2
\int_{-\infty}^{+\infty}\frac{d\theta}{2\pi}
\frac{1}{e^{mR \cosh\theta} -1} \,\,\,,
\label{bosonic}
\end{equation}
which is nothing else but the thermal one-point function
of $T_{\mu}^{\mu}$ for a free bosonic theory.

Let us consider now the Form Factors of $T_{\mu}^{\mu}$ associated
to the simplified model with kernel (\ref{delta}). In virtue of the
observed equivalence of this theory with a particular limit of the
Sinh--Gordon model, the Form Factors can obtained by a careful
$g \goto 0$ limit of the corresponding quantities of the Sinh-Gordon
model. They were computed in \cite{KM} and can be expressed as
\begin{equation}
\langle 0 | T_{\mu}^{\mu}(0) | \theta_1,\ldots,\theta_n\rangle =
\frac{2 \pi m^2}{F_{min}(i\pi)}
\left(\frac{4 \sin\pi\alpha}{F_{min}(i\pi)}\right)^{n-1}
{\cal Q}_n(x_1,\ldots,x_n) \prod_{i<j}
\frac{F_{min}(\theta_{ij})}{x_i+x_j}
\,\,\,.
\label{FFTheta}
\end{equation}
Few words on the above expression. The explicit form of
$F_{min}(\theta)$ can be found in \cite{KM}. For our purposes
we only need the functional equation satisfied by $F_{min}(\theta)$
\begin{equation}
F_{min}(\theta) F_{min}(\theta + i \pi) =
\frac{\sinh\theta}{\sinh\theta + i \sin\pi\alpha} \,\,\,.
\label{funeq}
\end{equation}
${\cal Q}_n$ is a symmetric polynomial in the variables
$x_i \equiv e^{\theta_i}$ given by
$$
{\cal Q}_n(x_1,\ldots,x_n) = \makebox{\rm det} \,M_{ij} \,\,\,,
$$
with the $(n-1) \times (n-1)$ matrix $M_{ij}$ given by
$$
M_{ij} = \sigma_{2i-j} [i-j+1] \,\,\,.
$$
In the above equation the symbol $[n]$ is defined by
$$
[n] \equiv \frac{\sin(n \alpha)}{\sin\alpha} \,\,\,,
$$
and $\sigma_k$ is the elementary symmetric polynomial given
by the generating function
$$
\prod_{i=1}^{n} (x+x_i) = \sum_{k=0}^n x^{n-k}
\sigma_k(x_1,x_2,\ldots,x_n) \,\,\,.
$$
In the limit $\alpha \goto 0$, the first polynomials ${\cal Q}_n$ are
given by
\begin{eqnarray}
{\cal Q}_2 & = & \sigma_1 \,\,\,;\nonumber \\
{\cal Q}_4 & = & \sigma_1 \sigma_2 \sigma_3 \,\,\,; \\
{\cal Q}_6 & = & \sigma_1 \sigma_5 \left[
\sigma_2 \sigma_3 \sigma_4 + 3 \sigma_3 \sigma_6 -
4 (\sigma_1 \sigma_2 \sigma_6 + \sigma_4 \sigma_5)\right]
\,\,\,.\nonumber
\end{eqnarray}

\subsection{Two--particle contribution}
By using eq.\,(\ref{FFTheta}), let us compute
\begin{equation}
\langle \theta_2,\theta_1 |T_{\mu}^{\mu} |\theta_1,\theta_2\rangle
=\lim_{\eta_1\goto 0} \lim_{\eta_2\goto 0}
\langle 0|T_{\mu}^{\mu} |\theta_1 + i\pi +\eta_1,\theta_2 +
i\pi +\eta_2,\theta_1, \theta_2\rangle \,\,\,.
\label{lim1}
\end{equation}
We will consider the contributions coming from the different terms
in (\ref{FFTheta}) separately.

By using the functional equation (\ref{funeq}), for the product of
$F_{min}(\theta_{ij})$ we have, in the above limit
\begin{equation}
\prod_{i<j}F_{min}(\theta_{ij}) \longrightarrow [F_{min}(i\pi)]^2
\frac{\sinh^2\theta_{12}}{\sinh^2\theta_{12} + \sin^2\pi\alpha}
\,\,\,.
\label{pfm}
\end{equation}
For the polynomial of the denominator we have
\begin{equation}
\prod_{i<j}(x_i+x_j) \longrightarrow A_1 A_2 \,x_1 x_2
(x_1 + x_2)^2 (x_1 - x_2)^2 \,\,\,,
\label{x1x2}
\end{equation}
where $A_k = (1-e^{i\eta_k}) \sim -i \eta_k$. Finally, for the polynomial
${\cal Q}_4$ in the numerator we obtain
\begin{equation}
{\cal Q}_4 \longrightarrow x_1 x_2 (x_1^2 + x_2^2)
\left[(A_1^2 + A_2^2) x_1 x_2 + A_1 A_2 (x_1^2 + x_2^2)\right]
\,\,\,.
\label{q4}
\end{equation}
We have now two possibilities. The first consists of keeping in the
above expression only the term multiplying the combination $A_1 A_2$ (and
disregarding those multiplying $(A_1^2 + A_2)$). This leads to the
computation of the connected Form Factor. In this case, combining
all terms and taking the limit (\ref{lim1}), we have
\begin{equation}
\langle \theta_2,\theta_1 |T_{\mu}^{\mu} | \theta_1,\theta_2
\rangle_{\rm conn}
= 4\pi m^2 \left(\frac{2 \sin\pi\alpha \cosh\theta_{12}}
{\sinh^2\theta_{12} + \sin^2\pi\alpha}\right)
\cosh\theta_{12}
\,\,\, .
\end{equation}
By taking now the limit $\alpha \goto 0$ and using
eq.\,(\ref{distribution}), we have
\begin{equation}
\langle \theta_2,\theta_1 |T_{\mu}^{\mu} | \theta_1,\theta_2
\rangle_{\rm conn}
= 4\pi m^2 \phi(\theta_1-\theta_2)
\cosh\theta_{12}
\,\,\, ,
\end{equation}
in agreement with eq.\,(\ref{2.18}).

The second possibility consists of taking the symmetric limit
considered by Delfino. This is obtained by taking
$A_1 = A_2$. In this case, the symmetric limit of
eq.\,(\ref{lim1}) produces
\begin{equation}
\langle \theta_2,\theta_1 |T_{\mu}^{\mu} | \theta_1,\theta_2
\rangle_{\rm sym}
= 8\pi m^2 \left(\frac{2 \sin\pi\alpha \cosh\theta_{12}}
{\sinh^2\theta_{12} + \sin^2\pi\alpha}\right)
\cosh^2\frac{\theta_{12}}{2}
\,\,\, .
\end{equation}
By taking now the limit $\alpha \goto 0$ and using
eq.\,(\ref{distribution}), we have
\begin{equation}
\langle \theta_2,\theta_1 |T_{\mu}^{\mu} | \theta_1,\theta_2
\rangle_{\rm sym}
= 8\pi m^2 \phi(\theta_1-\theta_2)
\cosh^2\frac{\theta_{12}}{2}
\,\,\, .
\end{equation}
Let us consider the expression (\ref{onedel}) up to the
the two--particle contribution. For the function $g(\theta,R)$
we have
\begin{equation}
g(\theta,R) = \frac{1}{1 - S(0) e^{-mR \cosh\theta}} =
\frac{1}{1+e^{- m R \cosh\theta}} \,\,\,,
\end{equation}
and then
\begin{equation}
\langle T^\mu_\mu \rangle_R - (T_\mu^\mu)_0 = 2\pi m^2
\int_{-\infty}^{+\infty}\frac{d\theta}{2\pi} \left[
\frac{e^{-m R \cosh\theta}}{1+ e^{-mR \cosh\theta}} +
2 \frac{e^{-2 m R \cosh\theta}}{(1+e^{-m R \cosh\theta})^2} +
\cdots \right]\,\,\,.
\label{primo}
\end{equation}
Expanding this expression in power of $e^{-m R \cosh\theta}$
up to $e^{-2 m R \cosh\theta}$ we have
\begin{equation}
\langle T^\mu_\mu \rangle_R - (T_\mu^\mu)_0 = 2\pi m^2
\int_{-\infty}^{+\infty}\frac{d\theta}{2\pi} \left[
e^{-m R \cosh\theta} + e^{-2 m R \cosh\theta} +
{\cal O}(e^{-3 m R \cosh\theta})
\right]\,\,\,.
\label{priexp}
\end{equation}
Comparing now this expression with eq.\,(\ref{therfac}), we explicitly
confirm the agreement found at this order by Delfino in his paper.

\subsection{Three--particle contribution}
By using eq.\,(\ref{FFTheta}), let us compute
\begin{equation}
\langle \theta_3,\theta_2,\theta_1 |T_{\mu}^{\mu}
|\theta_1,\theta_2,\theta_2\rangle
=\lim_{\eta_1\goto 0} \lim_{\eta_2\goto 0} \lim_{\eta_3\goto 0}
\langle 0|T_{\mu}^{\mu} |\theta_1 + i\pi +\eta_1,\theta_2 +
i\pi +\eta_2,\theta_3 + i\pi +\eta_3,\theta_1, \theta_2,
\theta_3\rangle \,\,\,.
\label{lim2}
\end{equation}
As before, let us consider the contributions coming from the different
terms separately. By using the functional equation (\ref{funeq}), for
the product of $F_{min}(\theta_{ij})$ we have, in the above limit
\begin{equation}
\prod_{i<j}F_{min}(\theta_{ij}) \longrightarrow [F_{min}(i\pi)]^3
\left(\frac{\sinh^2\theta_{12}}{\sinh^2\theta_{12} + \sin^2\pi\alpha}
\right)
\left(\frac{\sinh^2\theta_{13}}{\sinh^2\theta_{13} + \sin^2\pi\alpha}
\right)
\left(
\frac{\sinh^2\theta_{23}}{\sinh^2\theta_{23} + \sin^2\pi\alpha}
\right)\,\,\,.
\label{pfm3}
\end{equation}
For the polynomial of the denominator we have
\begin{eqnarray}
\prod_{i<j}(x_i+x_j) &\longrightarrow &A_1 A_2 A_3 x_1 x_2 x_3
\left[(x_1^2 - x_2^2) (x_1^2 - x_3^2) (x_2^2 - x_3^2)\right]^2 =
\nonumber \\
& & = 64 \,A_1 A_2 A_3 (x_1 x_2 x_3)^5 (\sinh\theta_{12} \sinh\theta_{13}
\sinh\theta_{23})^2
\,\,\,.
\label{x1x2x3}
\end{eqnarray}
For the polynomial ${\cal Q}_6$, we have two possibilities. The
first consists of keeping only the term multiplying the
combination $A_1 A_2 A_3$ (and disregarding all other expressions
which multiply the other monomials like $A_1^3$, $A_1^2 A_2$
etc.). This leads to the computation of the connected Form Factor.
In this case we have
\begin{eqnarray}
{\cal Q}_6^{\rm conn} &\longrightarrow & A_1 A_2 A_3 x_1 x_2 x_3
(x_1^2 + x_2^2) (x_1^2 + x_3^2) (x_2^2 + x_3^2) \times  \\
& &  \left[(x_1 x_2)^2 (x_1^2+x_2^2-2 x_3^2) +
(x_1 x_3)^2 (x_1^2 +x_3^2-2 x_2^2) + (x_2 x_3)^2 (x_2^2 + x_3^2 - 2 x_1^2)
\right] \nonumber
\end{eqnarray}
\label{q6conn}
and for the connected Form Factor, combining all terms, we obtain
\begin{eqnarray}
\langle \theta_3,\theta_2,\theta_1 |T_{\mu}^{\mu}
|\theta_1,\theta_2,\theta_2\rangle_{\rm conn} & = &
2 \pi m^2
\left(\frac{2 \sin\pi\alpha \cosh\theta_{12}}
{\sinh^2\theta_{12} + \sin^2\pi\alpha}\right)
\left(\frac{2 \sin\pi\alpha \cosh\theta_{23}}
{\sinh^2\theta_{23} + \sin^2\pi\alpha}\right) \times \nonumber \\
& & \times \frac{\sinh^2\theta_{13}}{\sinh^2\theta_{13} +
\sin^2\pi\alpha} \cosh\theta_{13}  + {\rm permutations}  \,\,\,.
\end{eqnarray}
By taking the limit $\alpha \goto 0$, we obtain the result
reported in formula (\ref{2.19}).

The second possibilities consists of considering the symmetric
limit, which is obtained by taking $A \equiv A_1=A_2=A_3$. Other
terms enter the polynomial in this case and we have
\begin{eqnarray}
{\cal Q}_6^{\rm sym} &\longrightarrow & A^3 x_1 x_2 x_3
(x_1 + x_2 + x_3) (x_1 + x_2) (x_1 + x_3 ) (x_2 + x_3)
(x_1 x_2 + x_1 x_3 + x_2 x_3) \times \nonumber \\
& &  \left[(x_1 x_2)^2 (x_1^2+x_2^2-2 x_3^2) +
(x_1 x_3)^2 (x_1^2+x_3^2-2 x_2^2) + (x_2 x_3)^2 (x_2^2 + x_3^2 - 2 x_1^2)
+ \nonumber \right.\\
& & -2 x_1 x_2 x_3 [x_2 (x_1-x_3)^2 + x_1 (x_2 - x_3)^2 + x_3 (x_1-x_2)^2]
\left. \right]
\end{eqnarray}
\label{q6sym}
Therefore, combining all the different contributions, in the symmetric
limit we have
\begin{eqnarray}
\langle \theta_3,\theta_2,\theta_1 |T_{\mu}^{\mu}
|\theta_1,\theta_2,\theta_3\rangle_{\rm symm} & = &
2 \pi m^2
\left(\frac{2 \sin\pi\alpha \cosh\theta_{12}}
{\sinh^2\theta_{12} + \sin^2\pi\alpha}\right)
\left(\frac{2 \sin\pi\alpha \cosh\theta_{13}}{
\sinh^2\theta_{13} + \sin^2\pi\alpha}\right) \times \nonumber \\
& & \times \frac{\sinh^2\theta_{23}}{\sinh^2\theta_{23} +
\sin^2\pi\alpha}  [2 (\cosh\theta_{12} + \cosh\theta_{13} +
\cosh\theta_{23}) + 3] \times \nonumber\\
& & \times \frac{\cosh\frac{\theta_{12}}{2}
\cosh\frac{\theta_{13}}{2}}
{\cosh\theta_{12} \cosh\theta_{13}}
\frac{(2 \cosh^2\frac{\theta_{23}}{2} -1)}
{\cosh\frac{\theta_{23}}{2}}
+
{\rm permutations} \nonumber \,\,\,.
\end{eqnarray}
By taking now the limit $\alpha \goto 0$ and using
eq.\,(\ref{distribution}), we obtain
\begin{eqnarray}
\langle \theta_3,\theta_2,\theta_1 |T_{\mu}^{\mu}
|\theta_1,\theta_2,\theta_2\rangle_{\rm sym} & = &
2 \pi m^2 \phi(\theta_1-\theta_2) \phi(\theta_1-\theta_3)
\,
\frac{
\cosh\frac{\theta_{12}}{2}
\cosh\frac{\theta_{13}}{2}}
{\cosh\theta_{12} \cosh\theta_{13}}
\times \nonumber \\
& & [2 (\cosh\theta_{12} + \cosh\theta_{13} +
\cosh\theta_{23}) + 3]
\frac{(2 \cosh^2\frac{\theta_{23}}{2} -1)}
{\cosh\frac{\theta_{23}}{2}}
\,
\nonumber \\
& & + {\rm permutations}  \,\,\,.
\end{eqnarray}
Once inserted into eq.\,(\ref{onedel}), we have
\begin{equation}
\langle T^\mu_\mu \rangle_R - (T_\mu^\mu)_0 = 2\pi m^2
\int_{-\infty}^{+\infty}\frac{d\theta}{2\pi} \left[
\frac{e^{-m R \cosh\theta}}{1+ e^{-mR \cosh\theta}} +
2 \frac{e^{-2 m R \cosh\theta}}{(1+e^{-m R \cosh\theta})^2} +
\frac{9}{2} \frac{e^{-3 m R \cosh\theta}}{(1+e^{-m R \cosh\theta})^3} +
\cdots \right]\,\,\, ,
\label{primosecondo}
\end{equation}
and by making an expansion up to $e^{-3 m R \cosh\theta}$ we have
\begin{equation}
\langle T^\mu_\mu \rangle_R - (T_\mu^\mu)_0 = 2\pi m^2
\int_{-\infty}^{+\infty}\frac{d\theta}{2\pi} \left[
e^{-m R \cosh\theta} + e^{-2 m R \cosh\theta} +\frac{3}{2}
e^{-3 m R \cosh\theta}
+ {\cal O}(e^{-4 m R})\right]\,\,\,,
\label{prisecexp}
\end{equation}
i.e. the third order coefficient disagrees with the corresponding
coefficient of eq.\,(\ref{geometrical}).

\resection{Conclusions}
In this paper we have critically analysed the status of the thermal
formalism for two--dimensional integrable field theory by comparing
the approach proposed by LeClair and Mussardo with the approach
proposed by Delfino. Whereas the first approach is able to reproduce
the one--point function of $T_{\mu}^{\mu}$ as given by the TBA, the
second one is in agreement with the TBA formula only up to the
two--particle contribution and differs otherwise. This has been
explicitly shown by considering a simple integrable model, where
all calculations can be performed analytically without relying on
the solution of integral equation. It would be useful to further
explore the subject and see whether or not the approach by
LeClair and Mussardo passes other tests.

\end{document}